\documentclass[a4paper]{revtex4}
\usepackage{graphicx}
\usepackage{fancyhdr}
\usepackage{amsmath}
\usepackage{color}

\newcommand{\Br}{{\mathrm{Br}}}

\begin{document}

\title{Galactic Center Gamma Ray Excess and Higgs Boson(s)}

%

\author{S. Ipek}
\affiliation{Department of Physics, University of Washington, Seattle, WA 98195, USA}

\begin{abstract}
There is evidence for an excess of gamma rays with $O({\rm GeV})$ energy coming from the Galactic Center in data from the Fermi Telescope. The spectrum of the excess is well fit by 30~GeV dark matter annihilating into a pair of $b$ quarks with a cross section expected from a thermal relic. For an explanation of this  excess, we study a renormalizable model where dark matter couples to the SM via a pseudoscalar that mixes with the $CP$-odd Higgs boson of a Two HIggs Doublet Model. We report the constraints on this model from direct detection, Higgs decays, and rare $B$ meson decays.
\end{abstract}

\maketitle

\thispagestyle{fancy}


\section{Introduction}
\label{sec:intro}
Eventhough dark matter (DM) accounts for more than a quarter of the energy density of the Universe, the only evidence of its existence comes from  gravitational observations such as galactic rotation curves, cluster mergers and the cosmic microwave background (see, e.g.~\cite{feng_dark_2010} and references therein). There are three ways to search for non-gravitational interactions of DM: (i) Collider searches where DM is produced through SM interactions, (ii) direct detection, and (iii) indirect detection where products of DM annihilation or decay are searched for.  

There is evidence for indirect detection of DM annihilaition, several groups finding an excess of gamma rays of energy $\sim$1-3~GeV in the region of the Galactic Center observed by the Fermi Gamma Ray Space Telescope~\cite{Daylan:2014rsa,Abazajian:2014fta}. Astrophysical sources are unlikely explanations for this excess due to its spectrum and morphology~\cite{Hooper:2013nhl}. On the other hand the spectrum of the excess has been fit by DM annihilating to a number of final states, depending on its mass, notably 10~GeV DM annihilating to $\tau^+\tau^-$ (and possibly other leptons) and 30~GeV DM to $b\bar b$ \cite{Barger:2010mc}. The size of the excess is compatible with an annihilation cross section roughly equal to that expected for a thermal relic, $\langle \sigma v_{\rm rel}\rangle=3\times10^{-26}~{\rm cm^3}/{\rm s}$, suggesting that it is actually the result of DM annihilation. 
Here we focus on a 30~GeV DM that annihlates to $b$ quarks. The effective pseudoscalar operator
\begin{align}
{\cal L}_{\rm eff}&=\frac{m_b}{\Lambda^3}\bar\chi i\gamma^5\chi \bar b i\gamma^5 b,
\label{eq:Leff}
\end{align}
where $\chi$ is the DM, has been the most studied interaction to explain the excess~\cite{Boehm:2014hva} since its direct detection cross section is velocity suppressed and spin dependent, rendering it safe from direct detection experiments. 

In the rest of this paper we summarize the results of~\cite{Ipek:2014gua}, as presented in \emph{Toyama International Workshop on Higgs as a Probe of New Physics 2015}, constructing a UV complete model that produces the pseudoscalar interaction in Eq.~\ref{eq:Leff} where DM-SM interactions are mediated by a pseudoscalar that mixes with the $CP$-odd Higgs boson of a Two HIggs Doublet Model (2HDM). We also report the constraints on this model.

\section{The Model}
\label{sec:model}
To construct a pseudoscalar-pseudoscalar interaction between DM and the SM, we expand the Higgs sector of the SM with a second Higgs doublet, which has enough degrees of freedom to mix with a pseudoscalar mediator. We take the DM to be a Dirac fermion, $\chi$, with mass $m_\chi$, coupled to a real, gauge singlet, pseudoscalar  mediator, $a_0$, through
\begin{align}
{\cal L}_{\rm dark}&=y_\chi a_0 \bar\chi i\gamma^5\chi.
\label{eq:dm-yuk-flavor}
\end{align}
The mediator couples to the SM via the Higgs portal in the scalar potential which is
\begin{align}
&V=V_{\rm 2HDM}+\frac{1}{2}m_{a_0}^2a_0^2+\frac{\lambda_a}{4}a_0^4+iBa_0H_1^\dagger H_2+{\rm h.c.}
\label{eq:Vport}
\end{align}
with $H_{1,2}$ the two Higgs doublets. We  assume that ${\cal L}_{\rm dark}$ and $V$ are CP-conserving 
(i.e. $B$ and $y_\chi$ are both real, and there is no CP violation in $V_{\rm 2HDM}$). 
In this case, $a_0$ does not develop a VEV. We write the most general CP-conserving 2HDM potential as
\begin{align}
V_{\rm 2HDM}=&\lambda_1\left(H_1^\dagger H_1-\frac{v_1^2}{2}\right)^2+\lambda_2\left(H_2^\dagger H_2-\frac{v_2^2}{2}\right)^2
+\lambda_3\left[\left(H_1^\dagger H_1-\frac{v_1^2}{2}\right)+\left(H_2^\dagger H_2-\frac{v_2^2}{2}\right)\right]^2
\\
&+\lambda_4\left[\left(H_1^\dagger H_1\right)\left(H_2^\dagger H_2\right)-\left(H_1^\dagger H_2\right)\left(H_2^\dagger H_1\right)\right]
+\lambda_5\left[{\rm Re}\left(H_1^\dagger H_2\right)-\frac{v_1 v_2}{2}\right]^2
+\lambda_6\left[{\rm Im}\left(H_1^\dagger H_2\right)\right]^2,
\nonumber
\end{align}
with all $\lambda_i$ real. We have also imposed a $Z_2$ symmetry under which 
$H_1\to H_1$ and $H_2\to -H_2$ to suppress flavor-changing neutral currents, which 
is only softly broken by $V_{\rm 2HDM}$ and $V_{\rm port}$.  
The potential is minimized at 
$\langle H_i\rangle=v_i/\sqrt2$, $i=1,2$, and the $W$ and $Z$ masses fix 
$v_1^2+v_2^2=v^2=(246~{\rm GeV})^2$. The angle $\beta$ is defined by $\tan\beta=v_2/v_1$. 
In unitary gauge we can decompose the doublets as
\begin{align}
H_i&=\frac{1}{\sqrt{2}}
\left(\begin{array}{c}
    \sqrt 2 \phi_i^+ \\ 
    v_i+\rho_i+i\eta_i \\ 
  \end{array}\right).
\end{align}
The spectrum contains a charged Higgs ($H^\pm$), two \emph{CP}-even Higgses ($H,h$), and two \emph{CP}-odd states ($A,a$) that are a mixture of the \emph{CP}-odd Higgs $A_0$ and the pseudoscalar mediator $a_0$,
\begin{align}
H^\pm=\sin\beta\,\phi_1^\pm-&\cos\beta\,\phi_2^\pm, \quad
H=\cos\alpha\,\rho_1+\sin\alpha\,\rho_2,\quad
h=-\sin\alpha\,\rho_1+\cos\alpha\,\rho_2,\notag\\
A&=\cos\theta\,A_0+\sin\theta\, a_0,\quad
a=-\sin\theta\,A_0+\cos\theta\,a_0,
\end{align}
where $A_0=\sin\beta\,\eta_1-\cos\beta\,\eta_2$. For details of the scalar sector, including definitions of the mixing angles, see~\cite{Ipek:2014gua}.

The DM coupling to the mediator in Eq.~(\ref{eq:dm-yuk-flavor}) is expressed in terms of 
CP-odd mass eigenstates,
\begin{align}
{\cal L}_{\rm dark}&=y_\chi \left(\cos\theta\,a+\sin\theta\,A\right)\bar\chi i\gamma^5\chi.
\label{eq:dm-yuk-mass}
\end{align}

We will work in a Type II 2HDM, where $H_2$ couples to \emph{up}-type quarks and $H_1$ couples \emph{down}-type quarks and leptons.  The couplings of the neutral scalar mass eigenstates are then rescaled from the SM Higgs values by
\begin{align}
\xi_{e,d}^h&=-\frac{\sin\alpha}{\cos\beta},~\xi_{e,d}^H=\frac{\cos\alpha}{\cos\beta},~\xi_{e,d}^A=\tan\beta\cos\theta,~\xi_{e,d}^a=-\tan\beta\sin\theta,
\\
\xi_u^h&=\frac{\cos\alpha}{\sin\beta},~~~~\xi_u^H=\frac{\sin\alpha}{\sin\beta},~~\xi_u^A=\cot\beta\cos\theta,~~\xi_u^a=-\cot\beta\sin\theta.
\end{align}

To simplify the analysis, we work close to the decoupling and alignment limit of the 2HDM where  $\alpha\simeq\beta-\pi/2$ and $m_h\ll m_H\simeq m_{H^\pm}\simeq m_{A_0}$. Since $h$ has SM-like couplings in this limit, we identify it with the 125 GeV Higgs boson.
\section{Dark Matter Annihilation}
\label{sec:ann}
For $m_a\ll m_A$, the dark matter annihilates to SM particles primarily through $s$-channel $a$ exchange. The velocity averaged annihilation cross section for $\chi\bar\chi\to{\rm SM}$ in 
the nonrelativistic limit is
\begin{align}
\langle \sigma v_{\rm rel}\rangle&=\frac{y_\chi^2s_{2\theta}^2 \tan^2\beta}{8\pi}\frac{m_\chi^2}{m_a^4}\left[\left(1-\frac{4m_\chi^2}{m_a^2}\right)^2+\frac{\Gamma_a^2}{m_a^2}\right]^{-1}\!\!\! \sum_{f=b,\tau,\dots}\!\!\!\!\! N_C \frac{m_f^2}{v^2}\sqrt{1-\frac{m_f^2}{m_\chi^2}}\simeq 3\times10^{-26}~\frac{\rm cm^3}{\rm s}\left(\frac{y_\chi \sin2\theta \tan\beta}{2.4}\right)^2,
\label{eq:annih}
\end{align}
for the experimentally favored DM mass of 30~GeV, taking $m_a=100~{\rm GeV}$ and ignoring
$\Gamma_a$.

It is possible to achieve values of the annihilation cross section 
compatible with the gamma ray excess and relic density constraints with modest values of 
the mixing angle $\theta$, provided $\tan\beta$ is somewhat large. At this value of 
$m_\chi$ and for $\tan\beta$ larger than a few, the $b\bar b$ final 
state accounts for about 90\% of the annihilation cross section with $\tau^+\tau^-$ 
making up nearly all the rest. This is in line with what is suggested by fits to the gamma ray excess.

\section{Constraints on the Dark Sector}
\label{sec:results}
\begin{figure}
\includegraphics[scale=0.6]{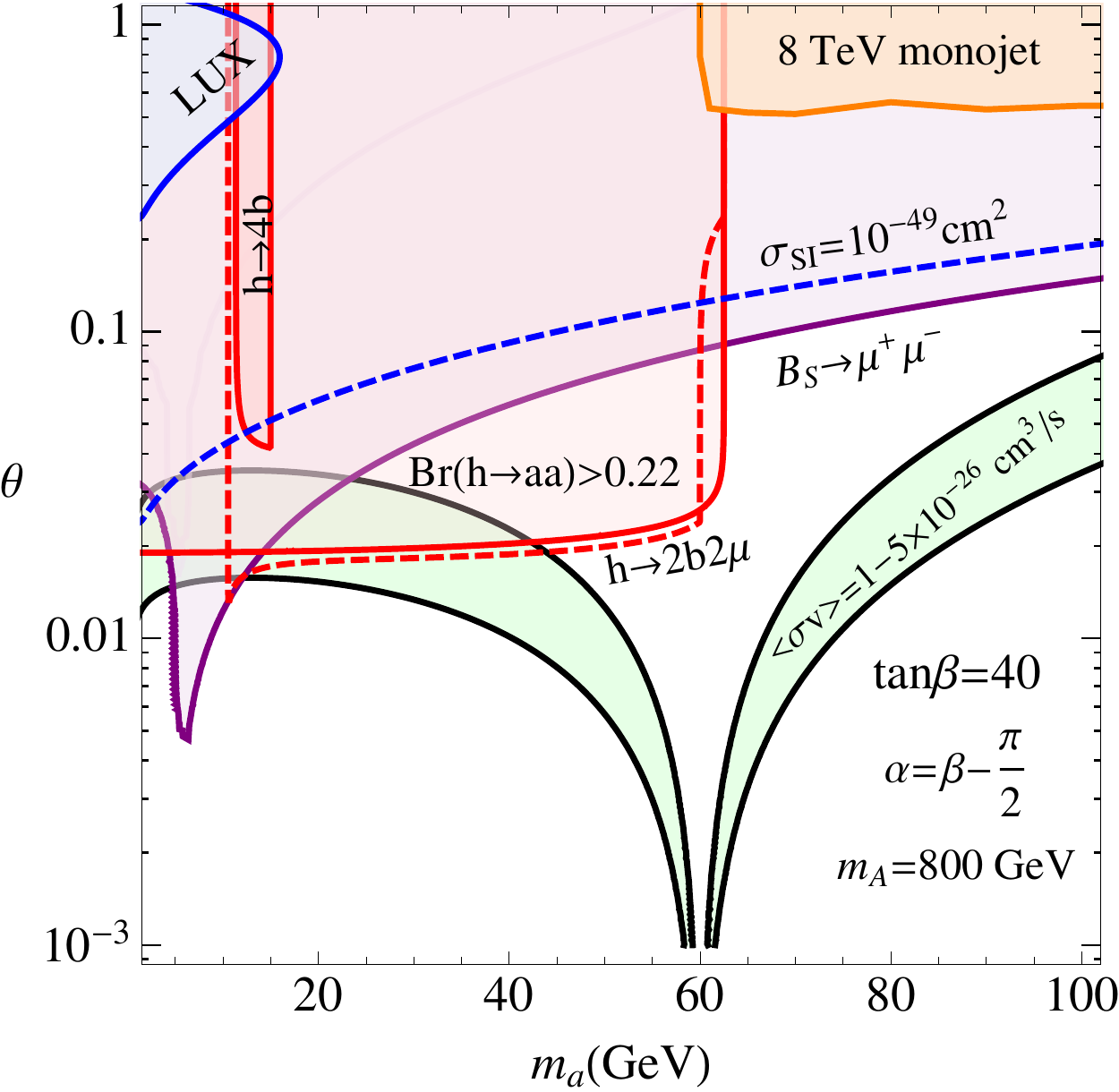}
\caption{Regions of mixing angle $\theta$ vs. $m_a$ that are ruled out or 
suggested by various measurements. We have fixed 
$m_{H,H^\pm}\simeq m_A=800~{\rm GeV}$, $\tan\beta=40$, $\alpha=\beta-\pi/2$, 
and $y_\chi=0.5$.
The area that gives an 
annihilation cross section
of $\langle \sigma v_{\rm rel}\rangle=1-5\times10^{-26}~{\rm cm}^3/{\rm s}$ 
as indicated by fits to the gamma ray excess is between the solid black lines 
(shaded in green).  
Other shaded regions are (projected) exclusion regions coming from various current (future) measurements. See text and also \cite{Ipek:2014gua} for details.
\label{fig:excl}}
\end{figure}

In this section we investigate the limits on the mediator mass and the mixing 
angle between the mediator and the pseudoscalar of the 2HDM.  We use the benchmark parameteres $m_H=m_{H^\pm}\simeq m_A=800~{\rm GeV}$, $\tan\beta=40$, $\alpha=\beta-\pi/2$, 
and $y_\chi=0.5$. We consider constraints  from direct detection, Higgs decays and rare $B$-meson decays. As discussed in~\cite{Ipek:2014gua}, this model is not well constrained by monojet searches due to the suppressed coupling to top at large $\tan\beta$. Our main results are summarized in Fig.~\ref{fig:excl}. 
\begin{enumerate}
\item \textbf{Direct Detection}

Pseudoscalar exchange between $\chi$ and quarks leads to (highly suppressed) spin-dependent scattering 
of the DM on nuclei. However, spin-independent interactions are generated at one-loop. For our benchmark values the dominant contribution to this comes from Higgs exchange triangle diagrams, leading to a spin-independent direct detection cross section of
\begin{align}
&\sigma_{\rm SI}\simeq 2.2\times10^{-49}~{\rm cm^2}\left(\frac{m_A}{800~{\rm GeV}}\right)^4\left(\frac{50~{\rm GeV}}{m_a}\right)^4
\left(\frac{m_\chi}{30~{\rm GeV}}\right)^2\left(\frac{\theta}{0.1}\right)^4\left(\frac{y_\chi}{0.5}\right)^4.
\end{align}

In Fig.~\ref{fig:excl}, we show the area of parameter space ruled out by the LUX limit of $8\times10^{-46}~{\rm cm^2}$~\cite{Akerib:2013tjd}. We also show the area that can be probed by a future cross section limit of $10^{-49}~{\rm cm^2}$.

\item \textbf{Higgs Decays}

If  $m_a<m_h/2$, Higgs can decay into a pair of $a$'s with the decay rate
\begin{align}
\Gamma\left({h\to aa}\right)&=\frac{\left(m_A^2-m_a^2\right)^2\sin^4 2\theta}{32\pi m_h v^2}\sqrt{1-\frac{4m_a^2}{m_h^2}}
\simeq840~{\rm MeV}\left(\frac{m_A}{800~{\rm GeV}}\right)^4\left(\frac{\theta}{0.1}\right)^4,
\end{align}
for $m_a\ll m_{h,A}$. This can impact LHC measurements of $h$, with a SM width of 4~MeV, which is SM-like to $10-20$\%.

For $m_a<m_h/2\simeq 2m_\chi$, the pseudoscalars will primarily go to $b$ quarks with a small branching to $\tau$ and $\mu$ pairs. The $h\to aa\to4b,2b2\mu$ signal will contribute to $h\to b\bar b$ searches~\cite{Curtin:2013fra}. A CMS search in $W/Z$-associated production at 7 and 8~TeV, 
$pp\to (W/Z)+(h\to b\bar b)$~\cite{Chatrchyan:2013zna}, sets a limit $\Br\left({h\to aa\to4b}\right)<0.7$ for $2m_b<m_a<15~{\rm GeV}$, and  $\Br\left({h\to aa\to 2b2\mu}\right)\lesssim 10^{-3}$ for $m_a>25~{\rm GeV}$. 

Since we are in the decoupling limit, the production cross section of the Higgs is unchanged from its SM value in this model. Therefore there are strong limits on unobserved final states, such as $aa$, that would dilute the signal strength in the observed channels. Given current data, this limits $\Br\left(h\to aa\right)<0.22$~\cite{Giardino:2013bma}.

We show in Fig.~\ref{fig:excl} the limits coming from  $\Br\left({h\to aa\to4b}\right)<0.7$ as well as the indirect constraint $\Br\left(h\to aa\right)<0.22$ and the limit that can be set by a future measurement of 
$\Br\left({h\to aa\to 2b2\mu}\right)<10^{-4}$. 

\item \textbf{B Physics Constraints}

A light $a$ can also potentially be constrained by its contributions to the decay 
$B_s\to\mu^+\mu^-$. For $m_a\ll m_Z$, the correction 
due to s-channel $a$ exchange can be simply written as~\cite{Skiba:1992mg}
\begin{align}
&\Br\left(B_s\to\mu^+\mu^-\right)\simeq\Br\left(B_s\to\mu^+\mu^-\right)_{\rm SM}
\times\left|1+\frac{m_b m_{B_s}t_\beta^2 s_\theta^2}{m_{B_s}^2-m_a^2}\frac{f\left(x_t,y_t,r\right)}{Y\left(x_t\right)}\right|^2,
\end{align}
with $x_t=m_t^2/m_W^2$, $y_t=m_t^2/m_{H^\pm}^2$, $r=m_{H^\pm}^2/m_W^2$, and $f$ and $Y$ are loop functions that are defined in \cite{Ipek:2014gua}.
The average of the LHCb and CMS measurements of this mode is
$\Br\left(B_s\to\mu^+\mu^-\right)=\left(2.9\pm0.7\right)\times10^{-9}$~\cite{CMS-PAS-BPH-13-007}. 
This should be compared against the SM prediction, which we take to be
$\left(3.65\pm0.23\right)\times10^{-9}$~\cite{Bobeth:2013uxa}. This offers a strong 
test of the model, especially for a light $a$, which we show in Fig.~\ref{fig:excl}.
\end{enumerate}
\section{Conclusions}
\label{sec:conc}
In order to explain the gamma ray excess from the Galactic Center, we have studied a 2HDM where the pseudoscalar mediator mixes with the CP-odd Higgs, giving rise to interactions between DM and the SM. 

A spin-independent cross section for direct detection arises at one-loop level in this model and is well below the current bound of $8\times10^{-46}{\rm cm}^2$ at a dark matter mass of 30~GeV.  We also consider decays of the 125~GeV SM-like Higgs boson involving the mediator. If the mediator is light $h\to aa\to 4b,\,2b2\mu$ can be constraining with data from the 14~TeV LHC. Additional contributions to $B_s\to\mu^+\mu^-$ in this model eliminate some of the favored parameter space for $m_a<10~{\rm GeV}$.

\begin{acknowledgements}
I would like to thank the organizers of \emph{Toyama International Workshop on Higgs as a Probe of New Physics 2015} for their hospitality. This work was supported in part by the U.S. Department of Energy under Grant No. DE-FG02-96ER40956. 
\end{acknowledgements}

\bibliography{ref2}

\end{document}